\documentclass[9pt,twocolumn,twoside]{opticajnl}
\journal{opticajournal} 

\setboolean{shortarticle}{true}

\title{Impact of Stimulated Brillouin Scattering on Modulation Instability in Fiber Fabry-Perot Resonators }

\author[1,*]{Alice Houard}
\author[2]{Thomas Bunel}
\author[1]{Matteo Conforti}
\author[1]{Saliya Coulibaly}
\author[1]{Arnaud Mussot}

\affil[1]{Laboratoire Physique des Lasers, Atomes et Molécules, UMR No. 8523, CNRS, University of Lille, 59000
Lille, France}
\affil[2]{Service OPERA-Photonique, Université libre de Bruxelles (ULB), Brussels, Belgium}

\affil[*]{alice.houard@univ-lille.fr}

\begin{abstract}
We report the experimental and numerical demonstration of stimulated Brillouin scattering–induced stabilization of Kerr-driven modulation instability in fiber Fabry--Perot resonators. We show that Brillouin scattering shapes the modulation instability spectrum into a stable multi--free-spectral-range cascade that is otherwise unstable in the absence of Brillouin effects. We further identify that this stabilization originates from the real part of the Brillouin response, leading to an asymmetric Kerr–Brillouin interaction governed by the relative position of the Brillouin frequency shift with respect to the cavity resonances. These results establish stimulated Brillouin scattering as an additional degree of freedom for controlling nonlinear dynamics in Kerr resonators.

\end{abstract}

\setboolean{displaycopyright}{false} 

\begin{document}

\maketitle

Optical frequency combs (OFCs) are powerful tools for precision distance metrology, spectroscopy, telecommunications, and microwave generation \cite{diddams2020opticalcomb, Fortier2019FrequencyCombReview}. Among the available platforms, microresonators have become a leading platform for Kerr-comb generation owing to their compactness, high quality factors, and strong nonlinearity \cite{pasquazi2018microcombs}. Fiber Fabry--Perot (FFP) resonators have recently emerged as a promising alternative \cite{Jia2020PhotonicFlywheel, Bunel2024THzSolitonComb, Obrzud2017TemporalSolitons, Li2024RamanSolitons}, offering competitive performance while enabling low-loss and straightforward integration into fiber systems, in contrast to the coupling constraints often encountered in microresonator-based devices.
FFP resonators exhibit specific nonlinear dynamics arising from their centimeter-scale length and their counter-propagating intra-cavity fields \cite{cole_theory_2018}. Their GHz-scale free spectral range (FSR), much larger than the $\sim$30 MHz Brillouin linewidth in silica fibers \cite{Kobyakov2010SBSReview} allows the $\sim$10 GHz Brillouin frequency shift to become resonant with a single cavity mode. As a result, the counter-propagating Brillouin Stokes waves can interact with the forward and backward Kerr fields, potentially modifying the comb-formation dynamics. 
Such Kerr--Brillouin interplay has already been shown to enable broadband switching-wave formation in single-mode FFP resonators under CW pumping \cite{bunel_brillouin-induced_2025}, ultra-stable cavity solitons through thermo-optic compensation \cite{Jia2020PhotonicFlywheel}, and narrow combs based on cascaded SBS processes in integrated Fabry--Pérot systems \cite{Buttner2014PhaseLockingFWM, Buttner2014CascadedSBS, Asano2016MicrobottleSBS}. Related Brillouin-assisted comb dynamics have also been reported in ring resonators, either integrated or fiber-based, in which the pump field propagates in a single direction \cite{Lucas2023BrillouinKerr, Huang2019BrillouinComb, Zhang2025BrillouinSolitons, Simon2026}.
However, although modulation instability (MI) is a fundamental mechanism for Kerr-comb formation in anomalous-dispersion resonators, the interaction between SBS and MI in FFP resonators remains largely unexplored. Apart from an early observation of Brillouin-enhanced hyperparametric oscillation in a fiber cavity \cite{braje2009brillouin}, little is known about how SBS affects the onset, modal organization, and stability of MI in this geometry.
Here, we investigate this interplay in the anomalous-dispersion regime of a FFP resonator and demonstrate that SBS can stabilize Kerr-driven modulation instability. We show numerically and experimentally that Brillouin scattering reshapes the MI spectrum into Brillouin-assisted cascades with a multi--FSR mode structure and enforces stable MI states that would be unstable in the absence of SBS. Furthermore, we analyze the frequency stability of the generated spectra and reveal an intrinsic asymmetry in the Kerr--Brillouin interaction arising from the real part of the Brillouin response.

\begin{figure}[ht!]
    \centering
    \includegraphics[width=0.9\linewidth]{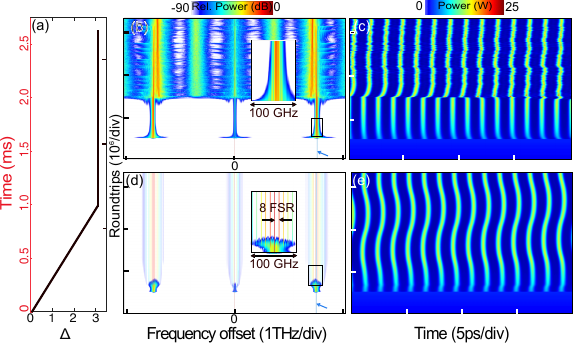}
    \caption{Numerical simulations. (a) detuning scan-stop applied to the pump. The right axis gives the round-trip number, matching the vertical axis of (b–e).  Spectral (b,d) and temporal (c,e) evolutions of the cavity output intensity during the scan-stop depicted in (a), without (b,c) and with (d,e) SBS. The blue line pointed by an arrow in (b) and (d) shows the prediction of the position of the MI side lobe at threshold from theory (~\cite{PhysRevA.109.013507}). The pump power is fixed at 137mW. Cavity parameters are given in Supplementary Table S1.
}
    \label{fig1}
\end{figure}
We first investigate the interaction between SBS and MI numerically. To this end, we integrate the mean-field model for FFP resonators derived in \cite{cole_theory_2018}, extended to include the Brillouin response following \cite{conforti_fast_2025}. Details of the mean-field model and the simulation parameters are provided in the Supplementary Material.
To generate the MI comb, the normalized pump detuning ( $\Delta = \delta/\alpha $) is scanned from $\Delta=0.3125 $ to $\Delta= 3.4375$ over $1$ ms, corresponding to approximately $N_{\mathrm{RT}}= 1.3\times  10^{6}$ cavity round trips, and is then fixed at its final value, as shown in [Fig.~\ref{fig1}(a)]. Pump power is fixed at 137mW.
To isolate the role of SBS, we perform two otherwise identical scan-stop simulations, showing the temporal and spectral evolutions at the cavity output, with the Brillouin gain switched off [Fig.~\ref{fig1}(b,c)] and on [Fig.~\ref{fig1}(d,e)].
 
In both cases, once the detuning reaches the MI threshold, characteristic symmetric sidebands emerge on both sides of the pump [Fig.~\ref{fig1}(b,d)], corresponding to the formation of a periodic temporal pattern [Fig.~\ref{fig1}(c,e)].  
The position of the MI sidebands is in excellent agreement with the theoretical prediction~\cite{PhysRevA.109.013507}, shown by the blue arrow in [Fig.~\ref{fig1}(b,d)], and very weakly affected by SBS (Supplementary Section S1).
When the SBS process is switched off [Fig.~\ref{fig1}(b,c)], the output spectrum evolves after about $1.5\times10^6$ round trips into a noisy MI state with additional interleaved sidebands appearing between the pump and the MI lobes. 
The unstable spectral dynamics lead to a periodic pattern with a significant timing jitter, round trip to round trip, as shown in Fig.~\ref{fig1}(c).
By contrast, when SBS process is on [Fig.~\ref{fig1}(d)], the MI sidebands are made of well defined and stable sub-comb of 8-FSR spacing (10.21 GHz), close to the SBS shift in silica (10.8 GHz), as we will detail below. In the time domain, this results in a stable periodic pulse train, with only a slow modulation that does not affect the short-term stability [Fig.~\ref{fig1}(e)].
These simulations show that SBS reorganizes an otherwise unstable MI regime into a stable spectral and temporal state.

\begin{figure}[ht!]
\centering\includegraphics[width=0.9\linewidth]{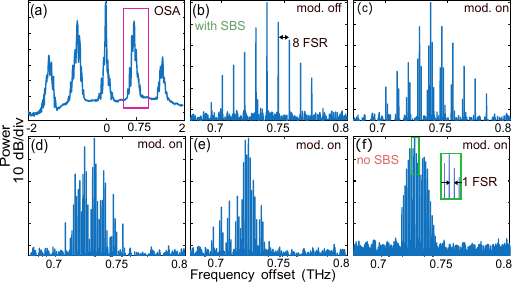}
\caption{Evolution of the MI spectrum with increasing phase modulation amplitude.
(a) Full optical spectrum recorded with an OSA.
(b–f) Zoom on the MI sideband centered at 750 GHz away from the pump measured with a BOSA for different RF powers applied to the PM: (b) phase modulation off, (c) 3 dBm, (d) 9 dBm, (e) 12 dBm, and (f) 25 dBm (respectively modulation amplitudes $\beta$ = 0.281, 0.396, 0.560, 3.53 rad). Increasing RF power progressively suppresses SBS.}\label{fig2}
\end{figure}

We experimentally verified these results using an 8.07-cm-long FFP cavity designed such that $N\times\mathrm{FSR}$ (N=8) lies close to the SBS frequency shift, enabling efficient Kerr-SBS interactions. 
The experimental setup is detailed in Fig.S1 of the Supplementary Material. We implemented a scan-stop experiment similar to the numerical one presented in Fig.~\ref{fig1}. 
Since SBS cannot be experimentally switched off in a FFP platform, its suppression was achieved by inserting a phase modulator right after the pump,  with the RF drive frequency set at the cavity FSR. The applied phase modulation redistributes the pump power among adjacent cavity modes and increases the effective SBS threshold~\cite{dianov_feasibility_1989}, allowing us to reduce
the effective SBS gain in the cavity.
We recorded the spectra of the cavity output using an OSA and a Brillouin optical spectrum analyzer (BOSA), the latter providing an ultra-high resolution (10 MHz) sufficient for resolving the cavity modes (around 1.28 GHz apart). The full optical spectrum recorded with the OSA is shown in Fig.~\ref{fig2}(a), where the pink box indicates the spectral region recorded by the BOSA, and displayed in Figs.~\ref{fig2}(b)–(f). In Fig.~\ref{fig2}(b), no RF power is applied to the PM, meaning that SBS is fully present.
We progressively increase the RF power in Figs.~\ref{fig2}(c)-(e) up to effective SBS suppression in Fig.~\ref{fig2}(f), corresponding to the RF power beyond which the spectrum no longer evolves. As we reduce SBS gain, the spectrum evolves from a Brillouin-MI cascade in which modes are separated by 8 FSRs (10.21 GHz, see  Fig.\ref{fig2} (b)) to a denser comb with modes spaced by the cavity FSR [inset in Fig.\ref{fig2} (f)]. This behavior is in close agreement with the dynamics predicted by numerics in Fig.~\ref{fig1}. Thus, by controlling the SBS gain through the phase modulator, we clearly evidence the transition from an SBS-organized MI spectrum to a dense multi-mode MI regime obtained when SBS is suppressed.

\begin{figure}[ht!]
    \centering
    \includegraphics[width=0.9\linewidth]{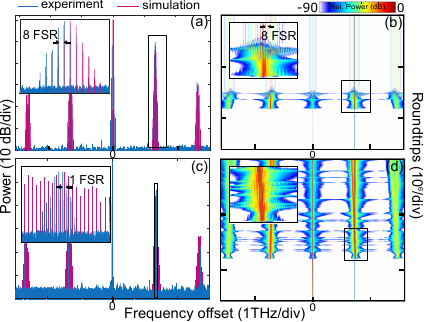}
    \caption{Comparison of experimental and simulated MI spectra with and without phase modulation. Left column: steady-state output spectra (blue: experiment, pink: numerics) obtained after the scan-stop procedure (a) without phase modulation and (c) with strong phase modulation suppressing SBS gain. Insets show a magnified view of the first MI sideband peak (black box). Right column: simulated spectral evolution during the scan-stop for the corresponding cases, (b) without and (d) with phase modulation.}
    \label{fig3}
\end{figure}

For a direct comparison with the experiments, we performed simulations that reproduce the actual phase-modulation protocol used to control SBS, rather than simply switching the SBS gain on and off as in Fig.~\ref{fig1}. These simulations therefore include both the Brillouin response and the pump phase modulation used experimentally to suppress SBS. 
Fig.~\ref{fig3}(a, c) show the steady-state spectra obtained at the end of the detuning scan-stop, while Fig.~\ref{fig3}(b, d) show the simulated round-trip-to-round-trip spectral evolution. 
Fig.~\ref{fig3}(a,b) corresponds to the case without phase modulation, i.e., with SBS present, whereas Fig.~\ref{fig3}(c,d) corresponds to the case with strong phase modulation, i.e., with SBS effectively suppressed.
Pink lines represent the simulated spectra whereas blue ones the experimental data. Insets show a magnified view of the first MI side lobe, outlined by the black box. Similarly to Fig.~\ref{fig1}, the spectrum stabilizes into a Brillouin-MI cascade when SBS is present whereas it remains unstable with multi-mode peaks structure when SBS is suppressed by phase modulation.
Agreement between simulation and experiment is excellent when SBS is present; when suppressed, simulated MI lobes are broader than experimental ones but retain the same one-FSR multi-mode structure.

\begin{figure}[ht!]
    \centering
    \includegraphics[width=0.9\linewidth]{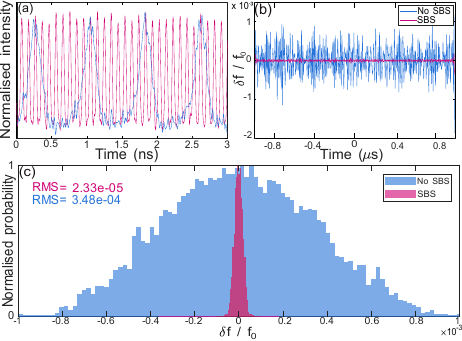}
    \caption{Experimental characterization of MI spectral stability with and without SBS. (a) Zoom on the time-domain beat signal obtained by optically filtering the cavity output of the first MI sideband, with  (pink) and without (blue) SBS. (b) Normalized instantaneous frequency fluctuations extracted from the beat signals, for both cases. (c) Normalized histograms of the instantaneous frequency fluctuations.}
    \label{fig4}
\end{figure}

Output spectra recorded by the OSA or BOSA provide averaged values, limiting the capture of the real time temporal evolution of the process. To demonstrate SBS-induced stabilization, we report round-trip-to-round-trip recordings of the cavity output field. 
Since we cannot resolve the fast-time evolution of the output field  directly (the MI frequency reaching several hundreds of GHz), we isolated one MI side lobe to capture the fast dynamics of the process (pink box in Fig. \ref{fig2}). 
To do so, we filtered out this side lobe at the cavity output, and the beating between the modes within this peak was recorded using a 70 GHz detection system (oscilloscope and photodetector). 
The resulting time-domain signals, acquired over several thousands of round-trips, are presented in Fig.~\ref{fig4}(a). When the PM is on (no SBS, blue trace), a periodic pulse train at 1.28~GHz (1×FSR) can be observed, while a 10.21~GHz (8×FSR) pulse train is observed when the PM is off (with SBS, pink trace). To quantify the timing jitter, the normalized fluctuations of the instantaneous frequency with respect to its mean value are shown in Fig.~\ref{fig4}(b), and the corresponding distributions are presented as histograms in Fig.~\ref{fig4}(c).
We computed these values by calculating the FFT of each signal; the first non-DC peak was isolated by filtering, and the IFFT was applied to retrieve the instantaneous phase, whose gradient gave the instantaneous frequency.
The instantaneous frequency fluctuations are markedly reduced, by around one order of magnitude, in the SBS case. Similar results were obtained when analyzing the first and second harmonics of the beat signal. These results further demonstrate the stabilizing effect of SBS on the multi-mode MI spectrum. 

\begin{figure*}[ht!]
\centering\includegraphics[width=0.9\textwidth]{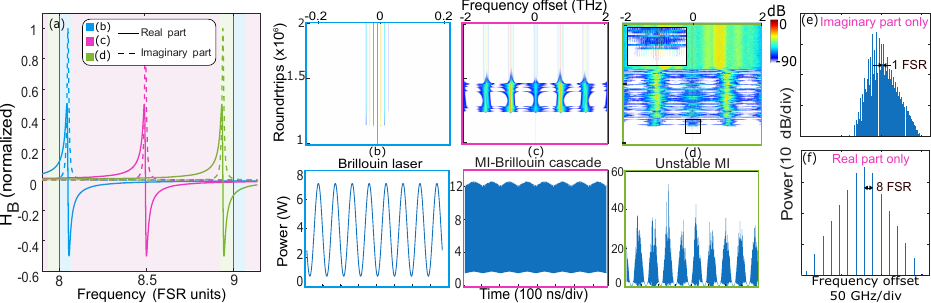 }
\caption{(a) Normalized Brillouin response functions as a function of shift position (in units of the cavity FSR) for SBS frequency shifts of 8.05, 8.5, and 8.95 FSR. The blue and green shaded regions correspond to conditions where the gain at an integer multiple of the cavity FSR exceeds the Brillouin lasing threshold.
(b–d) Typical spectral evolution (top) and final temporal profiles (bottom) obtained from identical scan-stop simulations for different SBS shift positions: (b) 8.05, (c) 8.5 and (d) 8.95 FSR.
(e,f) First MI sideband obtained when keeping only (e) the imaginary part and (f) the real part of the Brillouin response 
$H_B$.}
 \label{fig5}
\end{figure*}

To get a deeper understanding of the physical mechanism, we performed numerical simulations in which we varied the frequency shift of the SBS peak gain with respect to the cavity resonances. Although this situation is not experimentally accessible, we kept the cavity FSR fixed while tuning the SBS shift to provide a clear picture of the different dynamical regimes. In our model~\cite{bunel_brillouin-induced_2025, conforti_fast_2025}, $H_B$ denotes the complex Brillouin response function of the fiber, whose imaginary part corresponds to the Brillouin gain spectrum and whose real part describes the associated refractive index change. 
Fig.~\ref{fig5}(a) shows the normalized $H_B$ for several representative shift positions:  8.05$\times$FSR (blue curve, for which the SBS gain is slightly up-shifted with respect to the nearest cavity resonance),  8.5$\times$FSR (pink curve, experimental regime, for which the SBS gain lies far from cavity resonance) and 8.95$\times$FSR (green curve, for which SBS gain is slightly down-shifted compared to the nearest cavity resonance with the same offset as for 8.05$\times$FSR). Three qualitatively distinct regimes emerge, depending on this relative position. The corresponding simulations displaying the spectral and temporal behavior at the cavity output are shown in Figs.~\ref{fig5}~(b)–(d).

When the SBS shift is close to a cavity resonance (blue curve in Fig.\ref{fig5} (a)), the imaginary part of $H_B$, and therefore the Brillouin gain, is large at that mode frequency. In this region, the gain exceeds the Brillouin lasing threshold, causing the cavity to evolve into a pure cascaded Brillouin laser \cite{Hill1976MultiStokes}. The resulting spectrum consists of several laser lines spaced by the integer multiple of the cavity FSR that is closest to the SBS frequency shift [Fig.~\ref{fig5}(b), top]. In the time domain, this corresponds to a slow modulation of the intra-cavity field at 10.21~GHz (Fig.~\ref{fig5}(b), bottom). The MI process is not triggered, and no MI side lobes appear.
As the SBS shift is moved away from the cavity resonance, the Brillouin gain falls below threshold and MI can develop. In this intermediate regime (pink curve), the optical spectrum evolves into the stable Brillouin-MI cascade described experimentally [Fig.~\ref{fig5}(c), top]. In the time domain, the fast MI oscillations are modulated by a slower envelope associated with SBS [Fig.~\ref{fig5}(c), bottom], reproducing the 10.21~GHz modulation observed experimentally in Fig.~\ref{fig4}(a). This intermediate regime corresponds directly to the situation studied in Figs.~\ref{fig1}--\ref{fig4}, where the SBS shift used in both the experiments and the preceding simulations is 8.47$\times$FSR.

When the shift is brought back toward the next resonance (8.95$\times$FSR, green curve), its absolute detuning from resonance is similar to that of the 8.05$\times$FSR case (blue curve), but the dynamics are different. The inset in the top panel of Fig.~\ref{fig5}(d) shows that SBS sidebands appear around the pump before the onset of MI, producing the characteristic slow modulation of the intra-cavity field [Fig.~\ref{fig5}(d), bottom]. 
Contrary to the case in Fig.~\ref{fig5}(b), MI subsequently develops on top of this modulation, producing fast oscillations of the intra-cavity field together with a broadened and unstable MI spectrum [Fig.~\ref{fig5}(d), top]. 
This asymmetry cannot be explained by the Brillouin gain alone, since it is symmetric about the Brillouin resonance. Instead, it originates from the real part of $H_B$, which changes sign across the resonance. To isolate the respective contributions, simulations with an SBS shift equal to 8.5$\times$FSR were repeated while retaining only the imaginary or only the real part of $H_B$ [Figs.~\ref{fig5}(e,f)].
When the Brillouin response is limited to its imaginary part [Fig.~\ref{fig5}(e)], the MI side lobe contains modes separated by one FSR, and the stabilization into a Brillouin-MI cascade does not occur. 
In contrast, when only the real part of the Brillouin response is kept [Fig.~\ref{fig5}(f)], the MI lobe forms a clear Brillouin-MI cascade and the stabilization is recovered, confirming that the MI stabilization reported here is driven by the real part of the Brillouin response.

Although we detail only three representative shift values for clarity, we performed numerical simulations over a much wider range of SBS shifts. In every case, the cavity evolved into one of the three distinct regimes described above. When the SBS shift falls close to a cavity resonance on one side of the Brillouin response (blue-shaded region in Fig.~\ref{fig5}(a)), the system behaves as a cascaded Brillouin laser. When the SBS shift lies sufficiently far from any cavity mode (pink-shaded region), the real part of the Brillouin response stabilizes MI into the Brillouin-MI cascade observed experimentally. Approaching a cavity resonance from the opposite side (green-shaded region) produces an unstable MI state. These simulations show that the relative position of the SBS frequency shift with respect to the cavity resonances controls the MI--SBS interplay.

In conclusion, we demonstrated experimentally and numerically that SBS, rather than being a mere parasitic effect, can strongly modify the nonlinear dynamics in fiber Fabry--Perot resonators, leading to the formation of stable multi–FSR cascades instead of unstable modulation instability states observed in its absence.
Experimental measurements confirm this stabilization, showing a significant reduction of frequency fluctuations of inter-modal beat note within the MI side lobe when SBS is present. 
Numerical analysis shows that the system dynamics are governed by the relative position of the Brillouin frequency shift with respect to the cavity FSR, and reveals that the real part of the Brillouin response is responsible for the MI stabilization, controlling the emergence of stable or unstable modulation instability through Kerr–Brillouin interaction.
These results highlight SBS as a powerful degree of freedom for controlling nonlinear dynamics in fiber resonators and open new perspectives for engineering stable frequency comb states in these platforms.

\begin{backmatter}
\bmsection{Funding}
This project has received funding from the European Union’s Horizon Europe Research and Innovation Program under the Marie Sklodowska-Curie Grant Agreement No. 101169370. We acknowledge Agence Nationale de la Recherche (Programme Investissements d’Avenir) funding through project COMBY. The authors acknowledge the support of the CDP C2EMPI, and the French State under the France-2030 programme, the University of Lille, the Initiative of Excellence of the University of Lille, the European Metropolis of Lille.

\bmsection{Acknowledgment}
We thank Sarah Garzandat for fiber splicing assistance and Simon Bovinet for valuable discussions.

\bmsection{Disclosures}
The authors declare no conflicts of interest.

\bmsection{Data availability} 
Data underlying the results presented in this paper are not publicly available at this time but may be obtained from the authors upon reasonable request.

\bmsection{Supplemental document}
Supplementary material is available with this article.
\end{backmatter}

\bibliography{Biblio_paper_brillouin}

\bibliographyfullrefs{Biblio_paper_brillouin}


\ifthenelse{\equal{\journalref}{aop}}{%
\section*{Author Biographies}
\begingroup
\setlength\intextsep{0pt}
\begin{minipage}[t][6.3cm][t]{1.0\textwidth} 
  \begin{wrapfigure}{L}{0.25\textwidth}
    \includegraphics[width=0.25\textwidth]{john_smith.eps}
  \end{wrapfigure}
  \noindent
  {\bfseries John Smith} received his BSc (Mathematics) in 2000 from The University of Maryland. His research interests include lasers and optics.
\end{minipage}
\begin{minipage}{1.0\textwidth}
  \begin{wrapfigure}{L}{0.25\textwidth}
    \includegraphics[width=0.25\textwidth]{alice_smith.eps}
  \end{wrapfigure}
  \noindent
  {\bfseries Alice Smith} also received her BSc (Mathematics) in 2000 from The University of Maryland. Her research interests also include lasers and optics.
\end{minipage}
\endgroup
}{}

\end{document}